# Skyrmions in nanorings: a versatile platform for Skyrmionics

Dimitris Kechrakos[1‡], Vito Puliafito[2], Alejandro Riveros[3,4], Jiahao Liu[5,6,7], Wanjun Jiang[5,6], Mario Carpentieri[2], Riccardo Tomasello[2†], Giovanni Finocchio[4*]

[1]*Physics Laboratory, Department of Education, School of Pedagogical and Technological Education (ASPETE), Athens, Greece*
[2]*Department of Electrical and Information Engineering, Technical University of Bari, 70125 Bari, Italy*
[3]*Escuela de Ingeniería, Universidad Central de Chile, 8330601 Santiago, Chile*
[4]*Department of Mathematical and Computer Sciences, Physical Sciences and Earth Sciences, University of Messina, I-98166, Messina, Italy*
[5]*State Key Laboratory of Low-Dimensional Quantum Physics and Department of Physics, Tsinghua University, Beijing 100084, China*
[6]*Frontier Science Center for Quantum Information, Tsinghua University, Beijing 100084, China*
[7]*College of Advanced Interdisciplinary Studies & Hunan Provincial Key Laboratory of Novel Nano-Optoelectronic Information Materials and Devices, National University of Defense Technology, Changsha, Hunan, China, 410073*

[†] email: *riccardo.tomasello@poliba.it*
[‡] email: *dkehrakos@aspete.gr*
[*] email: *gfinocchio@unime.it*

## Abstract

The dynamical properties of skyrmions can be exploited to build devices with new functionalities. Here, we first investigate a skyrmion-based ring-shaped device by means of micromagnetic simulations and Thiele's equation. We subsequently show three applications scenarios: (1) a clock with tunable frequency that is biased with an electrical current having a radial spatial distribution, (2) an alternator, where the skyrmion circular motion driven by an engineered anisotropy gradient is converted into an electrical signal, and (3) an energy harvester, where the skyrmion motion driven by a thermal gradient is converted into an electrical signal, thus providing a heat recovery operation. We also show how to precisely tune the frequency and amplitude of the output electrical signals by varying material parameters, geometrical parameters, number and velocity of skyrmions, and we further prove the correct device functionality under realistic conditions given by room temperature and internal material defects. Our results open a new route for the realization of energy efficient nanoscale clocks, generators, and energy harvesters.

# I. INTRODUCTION

Magnetic skyrmions are fascinating topological textures with a number of intriguing fundamental properties and several potential engineering applications. Skyrmions are characterized by an integer topological index (skyrmion number) $Q = \frac{1}{4\pi}\int \boldsymbol{m} \cdot (\partial_x \boldsymbol{m} \times \partial_y \boldsymbol{m}) dx dy$ [1,2], which represents the number of times the magnetization vector wraps a surface of the unit sphere. Topologically, skyrmions transformation into a phase with a different $Q$ is forbidden from a mathematical point of view [1,2]. However, physically, such a "topological protection" provides an additional barrier for skyrmion annihilation/nucleation [3,4].

In view of skyrmion-based technological applications, two crucial requirements are necessary: electrical control and room temperature stability of skyrmions. So far, skyrmion manipulation by spin-orbit torque (SOT) is the most frequently studied approach [5–10]. However, promising alternative methods have been also proposed. Some of them are based on gradients of external magnetic field [11–14], perpendicular anisotropy [15,16], as well as thermal gradient (skyrmion-caloritronics) [17–19]. Room-temperature skyrmionic states have been observed in a wide range of materials, such as $B_{20}$ compounds [20,21], ferromagnetic single layer (FM) in contact with a heavy metal (HM) [10,22,23], $HM_1$/FM/$HM_2$ ferromagnetic multilayers [8,9,24–29], ferrimagnets [30,31], synthetic antiferromagnets [32–34], and combination of the previous ones [35–37].

A prime application of skyrmions is the racetrack memory that has stimulated a lot of research efforts [38–40]. In those devices, the digital information can be coded in the presence/absence of a skyrmion or in two different types of skyrmions [35,37,41]. However, skyrmion racetrack memory suffers from a big drawback, intrinsically linked to the topological character of ferromagnetic skyrmions, which is the occurrence of a finite skyrmion Hall angle [9,10,42] that quantifies the undesired drag of skyrmions in a direction that is transverse to the racetrack. Many strategies have been proposed to reduce and ultimately suppress the skyrmion Hall angle, from engineering anisotropy in ferromagnetic tracks [15,43,44] to the use of ferri [30]-and antiferro-magnets [45–48], including SAFs [49,50]. Predicted conventional applications also include spin-torque oscillators (STO) [51–54], microwave detectors [55], as well as logic gates [56–58]. Furthermore, skyrmions have been suggested for unconventional applications, such as in a reshuffle device for probabilistic computing [59,60], reservoir computing [61–63], true random number generators [64], Boolean logic computing gates [56,65–67], neuromorphic computing [68,69], artificial spiking neuron [70,71], memristive networks [72–74], and solution of the shortest path problem [75].

A promising geometry for skyrmion-based devices is a nanoring, where skyrmions have been previously studied experimentally with respect to their field gradient-driven motion [12], and theoretically for microwave emissions [76], and controlled nucleation [77]. Here, we expand the potential usage of such structure by designing three applications: (1) a clock with tunable frequency, (2) a skyrmion-alternator, and (3) a skyrmion-based energy harvester. We first demonstrate these functionalities by means of massive micromagnetic simulations and analytical calculations based on Thiele formalism. Subsequently, we propose their potential experimental implementations.

The working principle of the skyrmion clock (device-1) is based on the usage of a nanoring to generate a periodic electrical signal by repeatedly passing skyrmions through a detector region of the nanoring. Skyrmion motion is induced via standard SOT originated from a spin-current with a radial distribution. Skyrmion detection can be achieved via either the topological Hall resistivity [27,28,78] or the tunnelling magnetoresistance (TMR) [79,80]. These devices allow not only for frequency tunability via the number of skyrmions circulating in the nanoring, but also for a precise tunability on injected current and/or material parameters.

The skyrmion alternator (device-2) is realized with a contactless detection via a Faraday coil. The time variation of the stray field, generated by the periodic passage of the skyrmion under the Faraday coil, generates an electrical ac voltage with an amplitude proportional to the skyrmion velocity. As we demonstrate in the present work, for a set of parameters corresponding to state-of-the-art skyrmionic materials, the generated voltage amplitude can reach values of the order of a 1 µV. In such an application, to increase the energy conversion efficiency, we envisage a skyrmion motion driven by an engineered magnetic anisotropy [15,16], which can be considered as a ultralow power driving mechanism. The skyrmion alternator is an important prediction of our work since it represents the standard approach used in today's technology for the generation of electrical energy, which can be potentially used to drive nano/micro machines.

The skyrmion energy harvester (device-3) is based on the partial conversion of thermal energy due to heat dissipation into electrical energy. We envisage the presence of a heat source, for example, a microprocessor, located at the center of the nanoring. The thermal energy is dissipated towards the outer part of the nanoring thus creating a thermal gradient. The latter can trigger the skyrmion motion that is converted into an output voltage when the skyrmion passes underneath a Faraday coil. Considering also the feature of zero-input energy for gradient-driven motion, the suggested skyrmion application represents a very promising path towards the on-chip conversion from skyrmion motion and dissipated heat into electricity. We prove the stable operation of device-1 also under the influence of thermal fluctuations and internal material defects, and the robustness we demonstrated is also valid for the devices-2,3 scheme.

## II. DEVICE AND MODELING

### A. Device description

Figure 1(a) shows the proposed device for the skyrmion clock with tunable frequency. It is composed of a FM nanoring on top of an extended HM layer, with two gold contacts (yellow in the figure), one is inside the nanoring and the other surrounds the nanoring. The nanoring has external diameter $d_{nr}$, track width $w$ and a thickness $t_{FM}$. Typical material parameters are chosen to stabilize Néel skyrmions [39,81]. The driving force is a non-uniform radial charge current $j$ flowing into the HM that varies along the radial directions as $j_{HM}(r) = jR_0/r$, where $j$ is the injected current density and $R_0$ the inner radius of the nanoring. The current density generates an SOT with a counterclockwise (CCW) circular spin polarization **p** [39]. The skyrmion dynamics is characterized by a circular trajectory (see red curve in Fig. 1(a)) where the radial motion due to the skyrmion Hall angle [42] is almost compensated by the confining potential [39]. For this choice, the current-induced skyrmion Hall effect leads to a weakly inward trajectory of the skyrmions that causes their eventual annihilation at the inner boundary of the nanoring at large currents. The choice of an outward trajectory is less preferable, since the skyrmions would gradually enter regions of the nanoring with lower current density that would halt their motion. The device is equipped with a detector that is based on an MTJ having a width $w_d$=60 nm and a length equal to the width $w$ of the nanoring. Note that the detector can be also implemented with elliptical or circular geometry. The detector width $w_d$ is designed to be larger than the single skyrmion size considered (skyrmion diameter $D_{Sk} \approx$ 35-40nm) to achieve an efficient detection. The presence of a skyrmion in the detector region leads to variation of the magnetoresistive signal of the MTJ [79,80,82]. Moreover, this scheme allows for generating a clock signal in different positions of the nanotrack thanks to additional MTJ detectors and to the fact that the skyrmion is moving at the same velocity along the nanoring. In other words, we can achieve an intrinsic spatial synchronization of the clock frequency.

Figure 1(b) and (c) show the proposed device for the skyrmion alternator and energy harvester, respectively. Geometrically, these are similar to the previous one, but here the extended HM is necessary only to provide a sufficiently large interfacial Dzyaloshinskii-Moriya interaction (iDMI). The skyrmion motion is promoted by an engineered perpendicular anisotropy radial gradient (device-2) and a thermal radial gradient (device-3). The skyrmion motion in converted into an output electrical voltage $v_L$ when skyrmions pass below a Faraday coil (yellow circle in Fig. 1 (b) and (c)) located at height $h$ from the nanoring surface due to the well-known Faraday-Neumann-Lenz effect $v_L = -d\Phi/dt$, where $d\Phi$ represents the flux variation generated by the skyrmion movement below the coil. Table I summarizes the three proposed applications.

| Application | Source of skyrmion motion | Detection / Conversion |
|---|---|---|
| Skyrmion clock (device-1) | SOT with a tangential spin-polarization (**p**) | Magnetoresistive effect in a MTJ |
| Skyrmion alternator (device-2) | Linear gradient of the perpendicular anisotropy | Faraday coil |
| Energy harvester (device-3) | Linear thermal gradient | Faraday coil |

TABLE I. Summary of the versatile applications for skyrmions in nanoring.

### B. Micromagnetic model

The device design is based on massive micromagnetic simulations performed by means of the state-of-the-art micromagnetic solver *PETASPIN*, which numerically integrates the Landau-Lifshitz-Gilbert equation, augmented by the Slonczewski SOT term [83,84]:

$$\frac{d\mathbf{m}}{d\tau} = \mathbf{m} \times \mathbf{h}_{eff} + \alpha_G \left(\mathbf{m} \times \frac{d\mathbf{m}}{d\tau}\right) - \frac{g\mu_B \theta_{SH}}{2\gamma_0 e M_s^2 t_{FM}} [\mathbf{m} \times (\mathbf{m} \times (\hat{\mathbf{z}} \times \mathbf{j}_{HM}))] \qquad (1)$$

where $\mathbf{m} = \mathbf{M}/M_s$ is the normalized magnetization, $\tau = \gamma_0 M_s t$ is the dimensionless time, with $\gamma_0$ being the gyromagnetic ratio and $M_s$ the saturation magnetization. $\mathbf{h}_{eff}$ is the dimensionless effective field that includes the external, exchange, iDMI, magnetostatic and perpendicular anisotropy fields. $\alpha_G$ is the Gilbert damping, $g$ is the Landé factor, $\mu_B$ is the Bohr magneton, $\theta_{SH}$ is the spin-Hall angle, $e$ is the electron charge, $t_{FM}$ is the thickness of the ferromagnetic layer, $\hat{z}$ is the unit vector along the out-of-plane direction and $j_{HM}$ is the electrical current density flowing through the HM layer and giving rise to the SOT. The results presented here are for skyrmions in a nanoring with external diameter $d_{nr}$=460 nm, width $w$=150 nm and thickness $t_{FM}$=1 nm. Similar qualitative results are also observed for different nanoring size. We discretize the ring with cell dimensions 2×2×1 nm³. We use the following parameters, unless differently indicated [39], $M_S$ = 600 kA/m, exchange constant $A$ =10 pJ/m, iDMI parameter $D_m$ = 1.8 mJ/m², perpendicular anisotropy constant $K_u$ = 0.80 MJ/m³, spin-Hall angle $\theta_{SH}$ = 0.33 and $\alpha_G$ = 0.01.

### C. Thiele's equation

The analytical description of the skyrmion dynamics in the nanoring for the three device mechanisms can be achieved by the Thiele equation [39,85]

$$\boldsymbol{G} \times \boldsymbol{v} - \alpha_G D \cdot \boldsymbol{v} + \boldsymbol{F}_{SOT} + \boldsymbol{F}_K + \boldsymbol{F}_T = 0 \qquad (2)$$

where $\boldsymbol{G} = G\hat{z}$ is the gyrovector, $\boldsymbol{v}$ the skyrmion (core) velocity and $D$ the dissipative tensor matrix, respectively, while $\boldsymbol{F}_{SOT}$ is the force due to the SOT with tangential spin polarization (device-1), $\boldsymbol{F}_K$ the force due to the radial gradient of perpendicular anisotropy (device-2) and $\boldsymbol{F}_T$ the force due to the radial temperature gradient (device-3).

In particular, $\boldsymbol{F}_{SOT} = 4\pi B \boldsymbol{j}_{HM}$, where $\boldsymbol{j}_{HM}(r) = (jR_0/r)\hat{r}$ is the radially symmetric charge current density, $B = g\mu_B \theta_{SH}/2\gamma_0 eM_S^2 t_{FM}$ multiplied by a scaling factor to allow for racetrack shape effects [39], and $R_0$ the internal radius of the nanoring. On the other hand, the forces due to anisotropy and thermal gradient for devices-2,3 can be calculated through derivatives of the effective potential $V_{eff}$ associated with the effective field $\boldsymbol{h}_{eff}$ as obtained within the effective field theory [19]:

$$V_{eff} = 2\pi \left( 4A \left( \frac{\Delta}{R_{Sk}} + \frac{R_{Sk}}{\Delta} \right) + 4K \Delta - 2\pi D_m R_{Sk} \right), \qquad (3)$$

where $\Delta = \sqrt{A/K}$ is the domain wall width, $R_{SK} = \Delta/\sqrt{2\left(1 - \frac{\pi D_m}{4\sqrt{AK}}\right)}$ the equilibrium skyrmion radius, $K = K_u - \mu_0 M_S^2/2$ the effective magnetic anisotropy strength within the thin-film approximation and $D_m, K_u$ are the iDMI parameter and perpendicular anisotropy strength. Note that $V_{eff}$ is spatially uniform for device-1. Nevertheless, for device-2 it varies as $K_u = a_K r + b_K$ leading to a radially-dependent potential $V_{eff} = V_{eff}(r)$ which induces the force due to anisotropy gradient as

$$\boldsymbol{F}_K = -\frac{\hat{r}}{\mu_0 M_S^2} a_K \frac{\partial V_{eff}}{\partial K} = \hat{r} F_K(r), \qquad (4)$$

with $a_K$ the anisotropy gradient. Similarly, for device-3 we assume a temperature gradient in the radial direction, $T = a_T r + b_T$. We allow for temperature-dependent micromagnetic material parameters through the well-established scaling relations [19,81], $M_s \equiv M_s(T) = M_{s0}\left(1 - (T/T_{lim})^\delta\right)$, $A \equiv A(T) = A_0 \, m(T)^\alpha$, $D_m \equiv D_m(T) = D_{m0} \, m(T)^\beta$, and $K_u \equiv K_u(T) = K_{u0} \, m(T)^\gamma$, where $m(T) = M_s(T)/M_{s0}$ and $M_{s0}, A_0, D_{m0}, K_{u0}$ are the corresponding material parameters at zero temperature. The exponents are given as [19,81] $\alpha = \beta = \delta = 3/2$ and $\gamma = 3$ and $T_{lim} = 1120$ K is the Curie temperature. Eventually, the temperature gradient leads to a temperature-dependent effective potential $V_{eff} = V_{eff}(T)$, with a position-dependent temperature $T = T(r)$ and the resultant force due to the temperature gradient in device-3 reads:

$$\boldsymbol{F}_T = -\frac{\hat{r}}{\mu_0 M_s^2} a_T \frac{\partial V_{eff}}{\partial T} = \hat{r} F_T(r). \tag{5}$$

Since the three forces $\boldsymbol{F}_{SOT}$, $\boldsymbol{F}_K$ and $\boldsymbol{F}_T$ are all in the radial direction it is convenient to express Eq. (2) in polar coordinates $(r, \phi)$ of the skyrmion-core position, which can then be cast in matrix notation in the polar basis as

$$\begin{pmatrix} -\alpha_G D & -G \\ G & -\alpha_G D \end{pmatrix} \begin{pmatrix} \dot{r} \\ r\dot{\phi} \end{pmatrix} = -F(r) \begin{pmatrix} 1 \\ 0 \end{pmatrix} \tag{6}$$

where $F(r)$ denotes any of the three types of forces $F_{SOT}(r)$, $F_K(r)$, $F_T(r)$ acting on the skyrmion. From Eq.(6) we obtain for the radial and angular components of the skyrmion velocity:

$$\dot{r} = \alpha_G D\, F(r)/(G^2 + \alpha_G^2 D^2) \quad \text{and} \quad r\dot{\phi} = G\, F(r)/(G^2 + \alpha_G^2 D^2), \tag{7}$$

with the dot sign indicating a derivative with respect to dimensionless time $\tau$. Elimination of time and integration of Eq. (7) provides the skyrmion core trajectory

$$r(\phi) = r_0 \exp[\alpha_G D(\phi - \phi_0)/G], \tag{8}$$

which is a logarithmic spiral for the skyrmion motion in the three devices and $r_0$, $\phi_0$ are the initial values ($\tau=0$). Notice that the shape of the skyrmion trajectory, Eq.(8), is independent of the particular functional form of the radial force $F(r)$. On the other hand, the time evolution of the radial and azimuthal components of the skyrmion trajectory depend explicitly on the functional form of $F(r)$, which can be obtained by integrating Eq. (7). Analytical expressions of $F(r)$ were obtained from Eq.(4) and Eq.(5) using a symbolic algebra software.

### III. NUMERICAL RESULTS - Skyrmion clock (Device-1)

We discuss first our proposal for a skyrmion clock, where the motion is driven by the SOT from the electrical current and the skyrmion detection is performed via an MTJ detector with perpendicular pinned layer in order to sense, through a change of its resistance, the variation of the out-of-plane spatially-averaged magnetization component $<m_z>$. In the case of a single skyrmion circulating in the nanoring, the micromagnetic results are summarized in Fig. 2. Clearly, the time evolution of $<m_z>$ under the MTJ is characterized by a series of periodic dips ("comb" structure, Figs. 3(a)-(d)), which occur when the skyrmion passes underneath the detector. Simultaneously to the drop of magnetization, we show that the topological charge in the detector region reaches the value $Q = -1$, signifying the presence of a skyrmion under the MTJ.

The most important feature here is an evident periodicity of the skyrmion signal for up to 4-5 laps around the nanoring in a time interval up to approximately 200 ns that points to efficient pulse generation with constant frequency. However, we observe two features linked to the interaction of the skyrmion with the internal boundary of the nanoring as it completes more laps: (*i*) a delay in the last peak of the $<m_z>$ signal indicating the slow-down of the skyrmion, and (*ii*) smaller $<m_z>$ variations below the MTJ indicating a skyrmion size reduction, while $Q$ remains the same.

From recording the time-instants corresponding to the dips of the $<m_z>$ signal, we show that successive skyrmion laps by the detector vary linearly with time in Fig. 2(e). To support the results from micromagnetic simulations, we compare also with the analytical computations with the Thiele's equation described in previous section. For the clock device, $F(r) = F_{SOT}(r) = 4\pi B\, jR_0/r$ and Eq. (7) can be written as

$$\dot{r} = u\, \alpha_G D/Gr \quad \text{and} \quad r\dot{\phi} = u/r, \tag{9}$$

with $u = [4\pi BG/(G^2 + \alpha_G^2 D^2)]R_0 j$. Then the time evolution of the radial and azimuthal components read

$$r(\tau) = \sqrt{r_0^2 + 2u\alpha_G D\tau/G} \quad \text{and} \quad \phi(\tau) = \phi_0 + \frac{G}{2\alpha_G D}\ln\left(1 + \frac{2u\alpha_G D}{Gr_0^2}\tau\right) \tag{10}$$

The elapsed time $\tau_n$ for a skyrmion to perform $n$-laps through the detector is obtained from Eq. (10) by setting $\phi(\tau_n) = \phi_0 \pm 2\pi n$, as:

$$\tau_n = \frac{G\, r_0^2}{2\alpha_G Du}\left(e^{\pm 4\pi n\alpha_G D/G} - 1\right) \tag{11}$$

with the upper (lower) sign in the exponential term corresponding to CCW (CW) motion of the skyrmion along the spiral trajectory. The frequency (in Hz) of the $n$-th full circle along the spiral is defined as $f_n = (\tau_n - \tau_{n-1})^{-1}$ and becomes:

$$f_n = \gamma_0 M_s \frac{2\alpha_G Du}{Gr_0^2} e^{\mp 4\pi n\alpha_G D/G}\left(1 - e^{\mp 4\pi \alpha_G D/G}\right)^{-1} \tag{12}$$

Noted that the ratio of frequencies for successive laps is given by

$$\frac{f_{n+1}}{f_n} = e^{\mp 4\pi \alpha_G D/G} \tag{13}$$

which implies that the frequency of the generated electrical pulses due to sequential passages of the skyrmion through the detector region, is not constant. However, for most ferromagnetic materials of interest $\alpha_G D/G \ll 1$ and thus, nearly constant frequency can be achieved as indicated by Eq.(13). As shown in Fig. 2(e) and (f), the analytical and numerical outcomes are in a good agreement for both the skyrmion frequency and its number of laps in the detector region for different current densities and magnetic parameters, where the number of skyrmion laps and frequency predicted by the Thiele formalism were obtained by Eqs. (11,12), respectively. A time delay of the laps relative to the predictions of the analytical model is attributed to the skyrmion-boundary interaction that hinders the skyrmion motion. Interestingly, the boundary-free motion under a radial current density predicts a weakly accelerated skyrmion motion (see Fig. 4, in next section B), but the skyrmion-boundary interaction competes with this effect, restoring eventually a constant velocity (see Fig. 2(e)).

An advantage of this device is its frequency tunability which can be achieved by changing the driving current density, as indicated by the linear scaling of the frequency with current density in Fig. 2(e), and the number of skyrmions as discussed in the next paragraph. In addition, the material parameters can be used to set the desired frequency region [39] [81] (Fig. 2(f)). We also wish to highlight that the skyrmion configuration has an intrinsic reset mechanism due to the magnetostatic interactions which re-align the skyrmion position to the center of the nanoring in absence of applied current.

The thickness of the ferromagnet can influence different parameters. The most sensitive are the interfacial perpendicular anisotropy, the iDMI and the amplitude of the SOT. All of the three contributions scale linearly as a function of the thickness [86, 87]. The first two parameters affect the skyrmion stability and, above a certain critical thickness, the skyrmion is no longer stable. In addition, the reduction of the SOT with thickness, for a fixed set of parameters, gives rise to a reduction of the slope of the velocity-current curve. We have performed simulations with different values of the ferromagnetic film thickness (from 1.0 nm up to 2.0 nm) keeping a fixed set of parameters (see Section B) and constant current density. Our simulations indicated a drop of frequency up to 50% when the layer thickness increases from 1.0nm to 2.0 nm (not shown here).

### III.A. Clock with multiple skyrmions in the nanoring

Higher pulse frequencies can be achieved when an N-skyrmion ($N_{Sk}$) chain is stabilized in the nanoring. In simulations, we place multiple skyrmions in the nanoring and relax the system prior to applying the electrical current. The skyrmions are located at symmetrical points in the device. In particular, for the case of $N_{Sk}$=2, we create the skyrmions on a circle in the half-width of the nanoring at angular positions of 0º and 180º with respect to the detector region and, for the case of $N_{Sk}$=4, we place skyrmions at angular positions of 0º ,90º,180º and 270º. Figures 3(a) and (b) show the TMR signal due to the passage of multiple skyrmions for a fixed $j$=20 MA/cm$^2$ for $N_{Sk}$=2 and $N_{Sk}$=4, respectively, where the decrease of the $<m_z>$ period with the number of skyrmions is evident. To support this finding, we depict in Fig. 3(c) the number of laps *vs.* time for different current densities, where the curve slopes for $N_{Sk}$=4 is always larger than the $N_{Sk}$=2 case. Eventually, the previous results are confirmed in Fig. 3(d), where the frequency as a function of the current density for different $N_{Sk}$ is shown, thus underlying the role of the number of skyrmions in tuning the working frequency of device.

The Thiele's equation can be extended to account for $N_{Sk}$ skyrmions. In particular, Eq. (12) is also valid in the case of a skyrmion chain composed of $N_{Sk}$ equidistant and non-interacting skyrmions circulating on the nanoring. In this case, the detected frequency corresponds to the frequency of a single skyrmion scaled by the number of skyrmions in the chain

$$f_n(N_{sk}) = N_{sk} \cdot f_n(1) \qquad (14)$$

However, for a dense skyrmion chain, namely when the core-to-core distance approaches the skyrmion diameter, the detected frequency is anticipated to deviate slightly from the predictions of Eq.(14). The underlying physical mechanism causing these deviations is due to the skyrmion-skyrmion repulsive forces that alter the skyrmion shape and hence the dynamics of the interacting skyrmion chain. Nonetheless, for our study, the analytical predictions are in line with the micromagnetic simulations results. Weak deviations from the constant pulse frequency can be observed for high current density and long times (Fig. 3(c)) due to the interaction of the skyrmion with the inner boundary of the nanoring, however, due to the increased number of skyrmions in the device, a larger number of pulses has been generated relative to the single skyrmion case, before the linearity is perturbed.

Noticeably, the overlap of different datasets in Fig. 3(c), as for example the case of 2 skyrmions under $j$=20 MA/cm$^2$ and the case of 4 skyrmions under $j$=10 MA/cm$^2$, demonstrates the equivalence of the two controlling parameters of the pulse frequency, namely the driving current density $j$ and the number of injected skyrmions $N_{Sk}$. We wish to stress again that an important requirement for achieving linear scale-up of the signal frequency is to have the skyrmions at equidistant in the device, as non-equidistant skyrmions would generate an electrical pulse with two or more constituent frequencies, for initial states where the skyrmions are not equidistant we observe a rearrangement of them to the equidistant configuration thanks to the magnetostatic field (not shown) [12].

### III.B. Extended results of the Thiele's equation

Figure 4 shows the results as predicted from the solution of Thiele's equation, Eq. (11) for a larger time interval and the material parameters mentioned earlier [39]. Interestingly, the time evolution of the laps number follows a weak logarithmic dependence, which suggests that intrinsically the frequency generated by the skyrmion rotation cannot be constant, unless a damping parameter close to zero is considered. However, for an initial time interval (~200 ns), the number of laps-time relation can be quite accurately approximated by a linear dependence, that in turn implies a constant frequency of motion which is a prerequisite for clock functionality.

### III.C. Effect of sample disorder

Typical FM/HM samples are polycrystalline and include internal defects, which are usually responsible for skyrmion pinning [8,88]. To study the effects of sample defects on the skyrmion circular motion, we divide the nanoring in regions of random shape and size (grains) implementing the Voronoi tessellation algorithm [88]. We attribute a random value to the perpendicular anisotropy $K_u$ of each grain with a dispersion of 5% around the mean value. In addition to the dispersion in anisotropy values, another important parameter in polycrystalline samples is the ratio of the mean grain size $\bar{D}_g$ to the skyrmion diameter $D_{Sk}$ [50], which in our simulations is set to $D_{Sk} = 30$ nm. Figure 5 shows the results for the mean value of the number of skyrmion laps as a function of time. The average is extracted with 5 different Voronoi maps with $\bar{D}_g$ equal to 10 nm and 60 nm, respectively. We fixed $j$=40 MA/cm$^2$ which gives the largest frequency of the output signal in the ideal sample (Fig.2(f)). Our results demonstrate that for both small and large grain size the effect of the internal defects is negligible, thus proving the robustness of this application to structural disorder.

### III.D. Effect of thermal fluctuations

Temperature effects are important factors introducing randomness in the dynamics of skyrmion motion [60,82,89]. The thermal effects are accounted in Eq. (1) via a stochastic field $\boldsymbol{h}_{\text{th}}$, that is applied at each computational cell and reads $\boldsymbol{h}_{th} = \frac{\chi}{M_s}\sqrt{2\alpha_G k_B T/\mu_0 \gamma_0 \Delta V M_s \Delta t}$, with $k_B$ being the Boltzmann constant, $\Delta V$ the volume of the computational cubic cell, $\Delta t$ the simulation time step, $T$ temperature of the sample, and $\chi$ a three-dimensional white Gaussian noise with zero mean and unit variance [90, 91]. Values of stochastic field in different cells are uncorrelated. Figure 6 compares the <$m_z$> for a single skyrmion in the nanoring at $T$=0 K with a typical signal at $T$=300 K. The latter is noisy due to random thermal fluctuations and a running-average method with a time window of 1ns was used to extract the smooth signal.

Figure 6(c) shows the number of skyrmion laps as a function of time. The frequency $\bar{f}$ at $T$=300 K has been averaged over 10 repetitions over the thermal disorder and the corresponding dispersion is illustrated by the horizontal error bars. The micromagnetic results predict an average frequency $\bar{f}$ =49 MHz at $T$=300 K that is more than twice the frequency $f$ =20 MHz at $T$=0 K. Examination of the magnetization snapshots showed that, in the early stage of the motion (up to ~20ns), the skyrmion develops an outward radial velocity that drives it to the outer boundary, where it continues the circular motion with a higher velocity. We attribute such a higher velocity to the synergy between the current-driven motion and the temperature-driven gyrotropic motion of the skyrmions that has been previously studied in nanodots [81]. We have also checked that the width of the nanoring does not modify this behavior by performing simulations in wider nanorings ($d_{nr}$=460 nm, $w_d$=190 nm), which produce similar frequency enhancement results. Therefore, not only our proposed application can work at room temperature, but also the performance in terms of achievable frequency is better.

## IV. NUMERICAL RESULTS – Skyrmion Alternator and Skyrmion Energy Harvester

The second part of the work is dedicated to the design of a skyrmion alternator and a skyrmion energy harvester. Both applications share the same geometry (Figs. 1(b) and (c)) and working principle for reading the signal, but different driving mechanisms for the skyrmion circulating motion. In particular, here it is driven by gradients in the parameters (anisotropy, temperature), and not by an electrical current.

### IV.A. Skyrmion Alternator

For the skyrmion alternator, we considered a free-energy input source related to the presence of an engineered linear perpendicular anisotropy gradient. Its effect in extended FM has been already analyzed experimentally and theoretically [15,16], thus making it a viable alternative to the electrical current. In particular, the nanoring-shaped FM can be designed to have a radial gradient of the $K_u$, (see Fig. 1(b) for more details). Based on our previous results [16], where skyrmions move mainly perpendicularly to the $K_u$ gradient direction, we expect the skyrmion to circulate in the nanoring, and, every time it passes underneath a Faraday coil, the time variation of the stray field flux $\Phi$ is converted into a voltage pulse at the terminal of the coil due to the well-known Faraday-Neumann-Lenz effect $v_L = -d\Phi/dt$. In other words, the skyrmion rotation gives rise to an ac voltage, similarly to what occurs in the majority of today's electrical generators, where the mechanical rotation of a rotor generates an ac voltage.

### IV.B. Skyrmion energy harvester

The skyrmion motion in the energy harvester is instead promoted by a linear temperature gradient (Fig. 1(c)). Thermal gradients have been already proved to be a reliable motion source for skyrmions [17,19]. In particular, previous theoretical results pointed out that thermal gradients can be taken into account via linear gradients of the magnetic parameters ($A$, $D_m$, $K_u$, $M_s$) computed with proper scaling relations with temperature [19] (see also paragraph II.C before Eq. (5)). The combinations of all the parameters gradients give rise to a skyrmion motion mostly perpendicular to the gradient direction. With this in mind, our idea is to build the skyrmion-based nanoring around a thermal source, such as a microprocessor (Fig. 1(c)). Then, the dissipated thermal energy propagates radially towards the outer boundary of the nanoring, and we assume that the heat propagation induces a linear temperature gradient (hotter in the inside and colder in the outside), which is expected to drive the circular motion of skyrmion. Similar to the case of skyrmion alternator, this motion can be converted into an ac voltage. In other words, the proposed skyrmion energy harvester can partially recover dissipated heat in the form of electrical energy. Here, we are considering a point source for the thermal gradient in order to have a radial distribution, however the concept is also applicable for arbitrary spatial distribution of thermal sources which generates, however, nonperiodic output signals across the coil.

### IV.C. Conversion of the skyrmion motion into an electrical signal

For both cases of the skyrmion alternator and the skyrmion harvester, we are only interested in the maximum skyrmion velocity for given material parameters, since it determines the value of the generated voltage pulse. Therefore, we do not consider the effect of thermal fluctuations or internal material defects, which we expect to modify quantitatively our results (see previous paragraphs III.C and III.D), e.g. a smaller skyrmion velocity, but not qualitatively.

To this aim, we developed a post-processing tool for the calculation of the voltage pulse $v_L = -d\Phi/dt$ generated by the skyrmion motion. As a first step, we performed micromagnetic simulations with the parameters as in Ref. [81]. For the $K_u$ gradient, we considered a minimum (maximum) value at the inner (outer) edge of the nanoring equal to 0.6 MJ/m$^3$ (0.8 MJ/m$^3$). For the thermal gradient, we considered a minimum (maximum) value at the outer (inner) edge of the nanoring equal to 100K (300K). In both cases, the skyrmion rotates in the nanoring with a velocity $v_{Sk} \approx$ 20 m/s. These simulations are needed to: (*i*) confirm a steady-state size of the skyrmion while moving, and (ii) use that value of the skyrmion velocity as a reference for the post-processing tool in order to predict how the generated voltage changes with the velocity around that value (see Fig. 7). We computed the spatial distribution of the stray field due to the skyrmion up to 50 nm away from the nanoring surface, and considered a Faraday coil with a diameter similar to the skyrmion one.

Figure 7(a) shows the change of the stray field flux through the Faraday coil located at height $h$=20 nm from the surface of the nanoring due to the passage of a skyrmion below it. The flux $\Phi$ becomes negative as the skyrmion enters the region below the coil, and symmetrically goes back to a positive value as the skyrmion goes out from the coil region. The negative flux value is due to the skyrmion polarity which is negative in this study (-*z*-direction). The corresponding nV voltage pulse is illustrated in Fig. 7(b). This simple numerical experiment proves the working principle of the proposed device as alternator and energy harvester.

The amplitude of the voltage can be linearly tuned by the skyrmion velocity (Fig. 7(c)) which is directly linked to the amplitude of the anisotropy and temperature gradients, and non-linearly decreases with the height of the coil from the nanoring surface, as shown in Fig. 7(d).

To obtain a steady-state ac voltage signal, we can symmetrically deploy multiple skyrmions. Indeed, Fig. 7(e) depicts the output voltage due to the circulating motion of 10 skyrmions.

In addition, a strategy to enhance the $v_L$ amplitude is to use $(HM_1/FM/HM_2)_n$ ferromagnetic multilayers [8,9,24–29]. The asymmetric FM/HM$_{1,2}$ interfaces enhance the iDMI leading to increased skyrmion stability. Hybrid (i.e. thickness-dependent) or homochiral (i.e. pure Néel) skyrmions extending through all FM layers can form in such a multilayer [92,93], as a trade-off among the anisotropy, magnetostatic and IDMI energies. A skyrmion extending in all layers produces larger changes in the magnetic flux, when it crosses the Faraday coil, leading to larger values of the voltage amplitude. Therefore, magnetic multilayers instead of a single FM/HM bilayer, could be used to enhance the amplitude of the voltage pulse. In our study, we considered the physical parameters of Ref. [81] and the same geometrical parameters as in Fig. 1 for the FM layers. Successive FM layers are separated from each other by a 1 nm non-magnetic layer. We relax a skyrmion as a function of the number of layers and we obtain a pure Néel skyrmion in all the cases. We compute the stray field and from this the $v_L$ amplitude. Figure 7(f) shows that $v_L$ linearly increases with the number of layers up to almost half a μV.

Eventually, we wish to mention here that we are considering only one wrap in the Faraday coil, but the output voltage increases linearly with the number of wraps. For instance, with 10 wraps, the voltage can be easily increased to more than 1 μV.

We have also compared the simulation results of the skyrmion velocity in the device-2 and device-3 with the developed Thiele formalism, which are given by Eq.(7), with $F(r) = F_K(r)$ for device-2 and with $F(r) = F_T(r)$ for device-3. For a skyrmion moving along the mid-circle of the circular track, $r = R_0 + w/2$ and the skyrmion velocity is calculated to be $v_{Sk}$ = 37 m/s for device-2 and $v_{Sk}$ = 58 m/s for device-3. These values are of the same order but higher than the simulation results ($v_{Sk} \approx$ 20 m/s). We attribute the enhanced velocity in the analytical model due to the absence of repulsive forces from the inner boundary of the nanoring when the skyrmion approaches. In particular, both gradient-induced forces $F_K(r)$ and $F_T(r)$ point in the inward radial direction being the parameters used and the edge force points in the outward radial direction. To take into account this effect, we have considered a fitting factor in the radial forces as $F(r) \rightarrow n_f F(r)$. The analytical calculations are in good agreement with the simulation results for $n_f \approx 0.5$ in device-2 and $n_f \approx 0.3$ in device-3 showing that the effect of the edge force has a key role in the skyrmion dynamics in those devices as already shown for racetrack memories [39].

## V. SUMMARY AND CONCLUSIONS

We have demonstrated the versatile potential usages of skyrmions in nanorings by combining numerical micromagnetic simulations and analytical calculations based on the Thiele's equation. In particular, we have proposed and examined three applications that rely on the conversion of the skyrmion circulating motion into an electrical signal:

(1) a skyrmion clock, where the skyrmion motion driven by a radially-flowing current generates periodic voltage pulses when the skyrmion pass below an MTJ. The frequency of the device can be tuned via current, material parameters, and number of skyrmions, reaching 200 MHz. This design allows for an intrinsic spatial synchronization of the clock frequency;

(2) a skyrmion alternator, where the skyrmion motion driven by an engineered anisotropy gradient (no electrical input) generates a voltage pulse close to the µV order due to the variation of the stray field through a Faraday coil. This idea is analogous to today's electrical generators, where the mechanical rotation of a rotor generates electrical voltage. We anticipate here that the gradient of other parameters can be used for the development of skyrmion alternators, such as magnetic field gradient and/or iDMI gradient;

(3) if the skyrmion motion is driven by thermal gradients, an energy harvester can be designed to partially recover the dissipated heat from a thermal source, such as microprocessor.

Here, we summarize that experimental studies of those nanoring geometries can use low-damping ferromagnetic materials, such as amorphous CoFeB [60]. Stacks hosting hybrid skyrmions can be designed in such a way that the skyrmion Hall angle is reduced to optimize the performance of the nanoring-based devices in order to increase the skyrmion velocity. This can reflect not only in faster clock, but also in an enhanced of the voltage generated by the skyrmion motion. In addition, we wish to stress that those skyrmion hosting materials can be integrated with MTJs stacks, as already demonstrated experimentally, making those device configurations more feasible [80].

At this stage of the research, it is not possible to perform a quantitative performance comparison between skyrmionic devices and existing clock solutions since the development of skyrmionic devices is still at the level of critical function or proof of concept establishment (Technological Readiness Level 3). From a qualitative point of view, the skyrmion-based clock implementation can be used to reduce the clock skew, enforcing the time synchronization between different parts of the circuits. In other words, the skyrmion being a soliton can travel with a reduced distortion and can be used locally for the generation of the clock timing. Our vision about the skyrmion alternator is the possibility to create a new generation of nano engines, hence the potential scalability at nanoscale will be the advantage as compared to current technology.

The same idea of scalability at nanoscale applies to the concept of energy harvesting. In this field, the current technology is facing several challenges and skyrmion based energy harvesting can give rise to a new direction in this research field [94].

**ACKNOWLEDGMENTS**

DK acknowledges financial support from the Special Account for Research of the School of Pedagogical and Technological Education through program "*Educational and Research Infrastructure Support*" (No 52922) and hospitality by the Politecnico di Bari during the course of this work. The research has been supported by the project PRIN 2020LWPKH7 funded by the Italian Ministry of Research. RT, VP, MC and GF are with the PETASPIN team and thank the support of PETASPIN association (www.petaspin.com).

AR acknowledges the kind hospitality in the University of Messina, during the course of this work and financial support from CIP2022036. The work carried out at Tsinghua University was supported by the Basic Science Center Project of National Natural Science Foundation of China (NSFC Grant No. 52388201), National Key R&D Program of China (Grant No. 2022YFA1405100), the NSFC distinguished Young Scholar program (Grant No. 12225409), the Beijing Natural Science Foundation (Grant No. Z190009), the NSFC (Grant Nos. 52271181, 51831005), the Tsinghua University Initiative Scientific Research Program and the Beijing Advanced Innovation Center for Future Chip (ICFC).

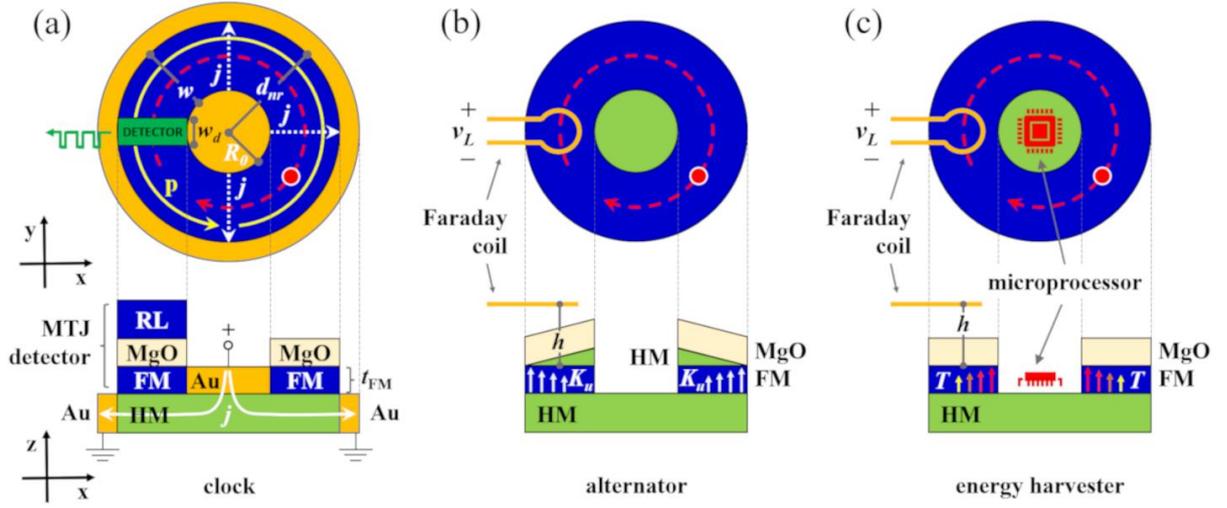

FIG. 1. Schematic representations of the skyrmion in a nanoring for the three applications envisaged in this work. (a) Clock with tunable frequency where the skyrmion circular motion is driven by an electrical current $j$ flowing between the internal and external Au electrodes via the HM (radial distribution of the current, circular distribution of the spin-polarization **p**). The skyrmion is detected via a localized MTJ, which has the FM as its free layer, and converts the variation of the out-of-plane magnetization into an electrical pulse. (b) Alternator where the skyrmion circular motion is driven by a perpendicular anisotropy gradient due to insertion of a HM wedged layer, as in Ref. [15]. The free-energy-driven motion of the skyrmion is converted into an electrical pulse $v_L$ by the Faraday coil located at a distance $h$ from the FM surface. The bottom HM is necessary only to reach a finite iDMI value. (c) Energy harvester where the skyrmion circular motion is promoted by a thermal gradient generated by a thermal source, e.g. a microprocessor, located in the middle of the nanoring. As in (b), the skyrmion motion is converted into an electrical pulse $v_L$ by the Faraday coil, and the bottom HM is necessary only to reach a finite iDMI value.

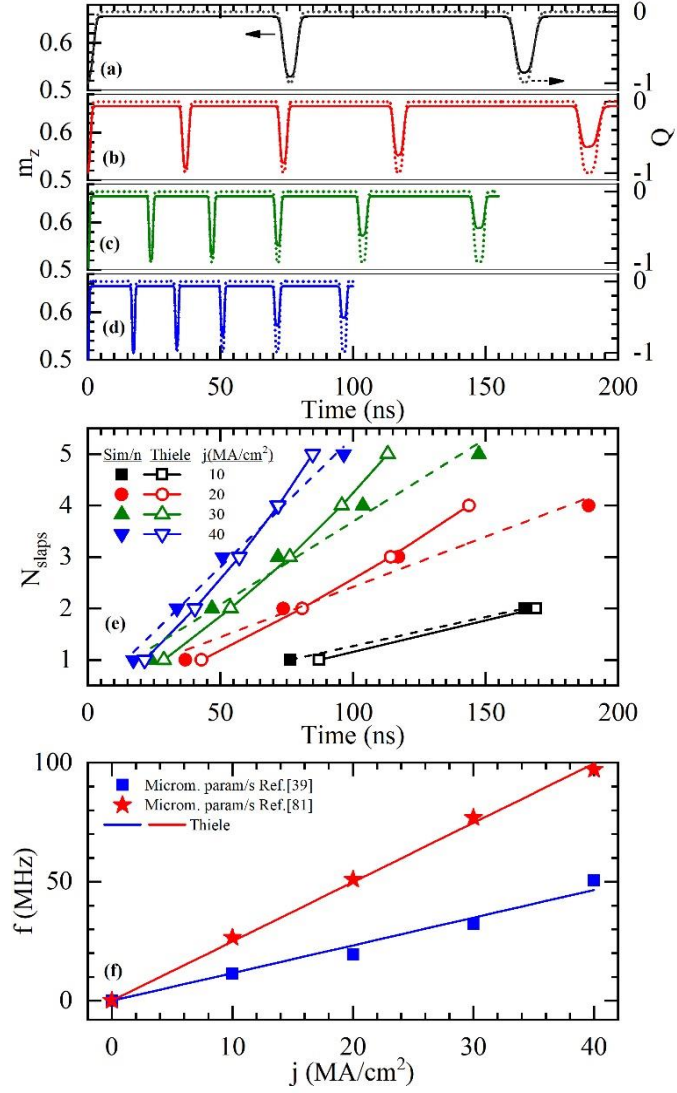

FIG. 2. Single skyrmion in a nanoring driven by an electrical current. (a-d) Time-evolution of the spatially-averaged out-of-plane component of the magnetization (solid lines) and the topological charge (dashed lines) in the detector region for different current densities $j$=10 (black), 20 (red), 30 (green), and 40 (blue) MA/cm$^2$, respectively. (e) Number of skyrmion laps in the detector region as a function of the elapsed time for various current densities, as indicated in the legend. A comparison of the analytical model predictions from Eq. (12) for $N_{sk}$=1 (open markers) with micromagnetic simulations results (closed markers) is shown. Dashed lines are linear fits of the micromagnetic simulation data for extracting the frequency of the skyrmion motion reported in (f). (f) Dependence of the output signal frequency on the electrical current density. Solid lines are the predictions of the analytical model, Eq. (12), while symbols refer to the micromagnetic simulations results.

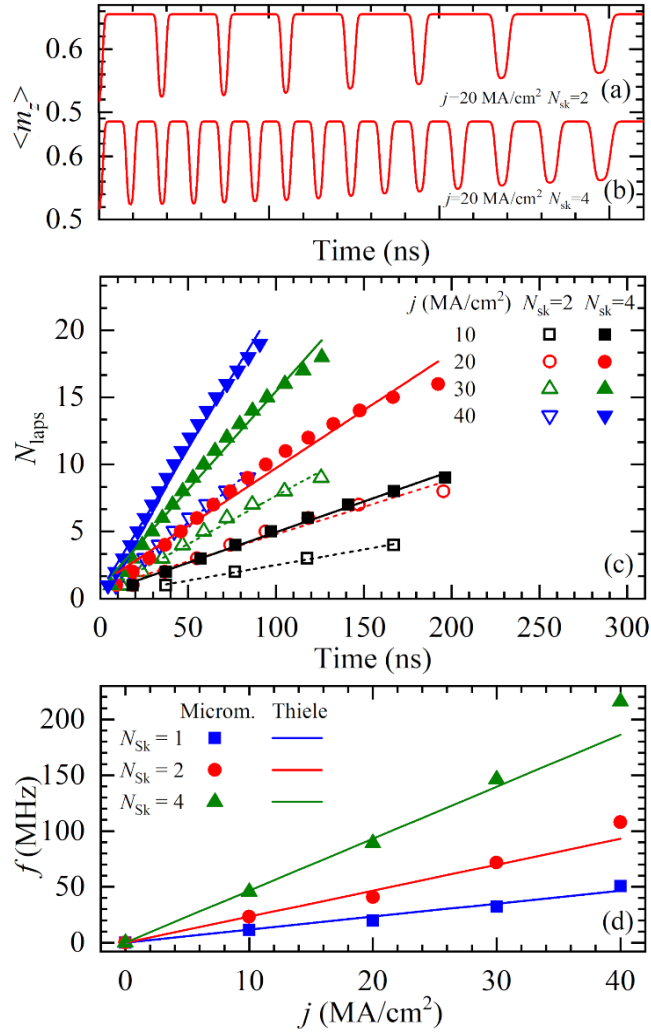

FIG. 3. Time-evolution of the spatially-averaged out-of-plane component of the magnetization $<m_z>$ for (a) two and (b) four skyrmions in the nanoring driven by current density $j$=20 MA/cm$^2$. (c) Number of laps in the MTJ detector as a function of the elapsed time for 2-skyrmion (open symbols) and 4-skyrmion (closed symbols) chains and various driving current densities. Straight lines (dashed for $N_{sk}$=2, solid for $N_{sk}$=4) are linear fits to the data used to extract the corresponding frequency values, which are shown as markers in (d). (d) Dependence of the output signal frequency on the electrical current density. Solid lines are the predictions of the analytical model, Eq. (14), while symbols refer to the micromagnetic simulations results.

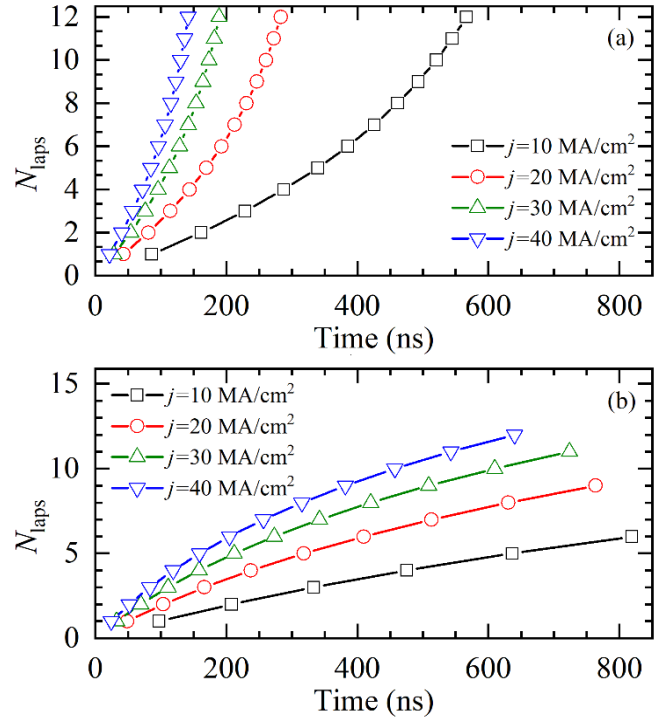

FIG. 4. Number of skyrmion laps in the MTJ detector region as a function of the elapsed time ($t_n$) given by Eq. (11), for a skyrmion circulating (a) inwards and (b) outwards. At t = 0 s, the skyrmion is located at the detector position in the middle of the ring width, that is $r_0 = R_0 + w/2$ and $\phi_0 = \pi$.

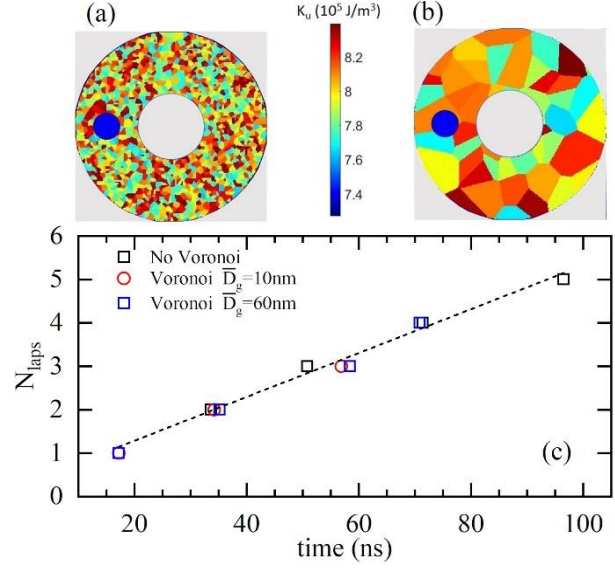

FIG. 5. Examples of Voronoi tessellation for (a) small grains with average size $\bar{D}_g = 10 nm$ and (b) large grains with average size $\bar{D}_g = 60 nm$. The blue disk on each nanoring is a sketch of a skyrmion with diameter $D_{sk} \approx 30 nm$. The colorbar indicates the values of the perpendicular anisotropy constant. It is characterized by 5% dispersion around the mean value $\bar{K}_u = 0.8\ MJ/m^3$. (c) Mean number of skyrmion laps as a function of time for a single skyrmion circulating in an ideal nanoring (black squares) and disordered nanorings (red circles, blue triangles). The applied current density is $j$=40 MA/cm$^2$ and the average grain size is $\bar{D}_g = 10 nm$ (red circles) and $\bar{D}_g = 60 nm$ (blue triangles). Error bars due to disorder are of the size of the marks. The straight dashed line is a linear fit to all data and serves as guide-to-eye.

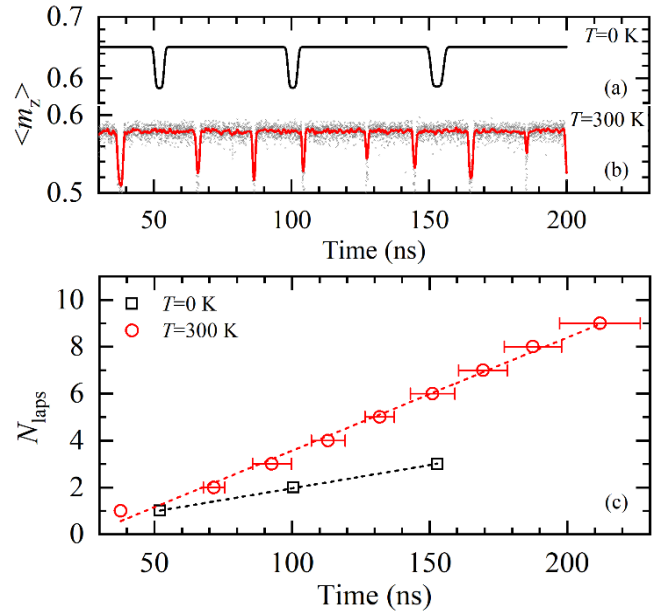

FIG. 6. (a-b) Time evolution of the out-of-plane component of the magnetization $<m_z>$ below the MTJ at $T=0$ K and $T=300$ K, for a single skyrmion in the nanoring under a driving current density $j=20$ MA/cm$^2$. An out-of-plane field ($H_z=300$ mT) and a slightly stronger DMI value ($D_m=4$ mJ/m$^2$) is used in order to stabilize the skyrmion at the elevated temperature ($T=300$ K). Gray markers in (b) indicate the raw noisy signal while the solid red line refers to the smooth signal, as obtained by a running-average method with a time window of 1ns. (c) Number of skyrmion laps in the detector region as a function of time at $T=0$ K (squares) and $T=300$ K (open red circles with corresponding error bars). Straight dashed lines are linear fits to data.

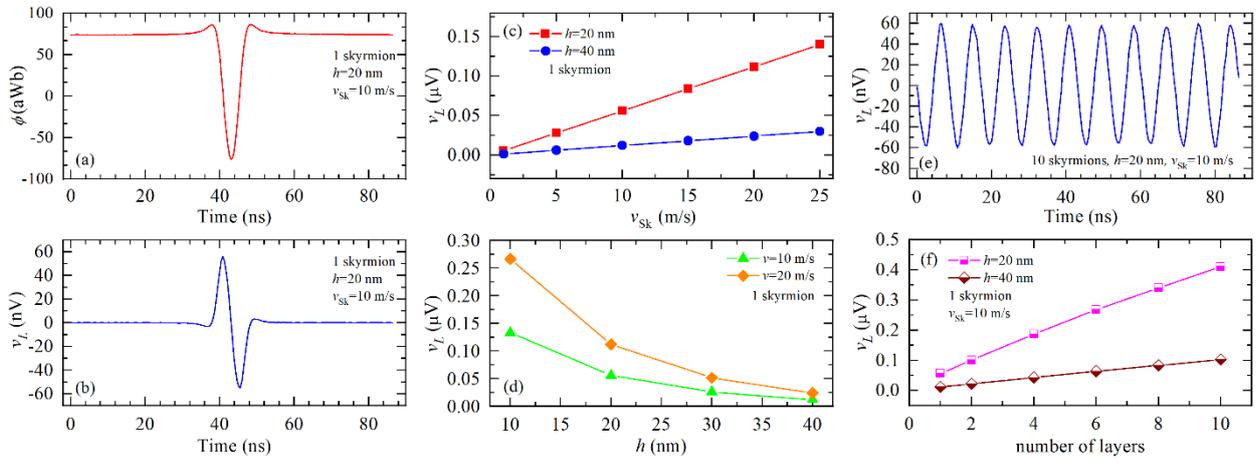

FIG. 7. AC voltage generated at the terminals of a Faraday coil due to skyrmion circulating motion. (a) Magnetostatic flux time-variation corresponding to a passage of 1 skyrmion underneath the coil for a skyrmion velocity $v_{Sk}$ = 10 m/s and a coil height $h$ = 20 nm. (b) Generated voltage pulse due to the flux variation in (a). (c) Amplitude of the voltage pulse as a function of the skyrmion velocity for two heights of the coil. (d) Amplitude of the voltage pulse as a function of the coil height for two skyrmion velocities. (e) Steady-state ac voltage output due to the consecutive passage of 10 skyrmions below the coil. (f) Amplitude of the voltage pulse as a function of the number of layers of the nanoring FM for two coil heights when $v_{Sk}$=10 m/s.